\documentclass[12pt]{iopart}

\usepackage{iopams}
\usepackage{graphicx}
\begin{document}

\title{Experimental observation of entanglement duality for identical
particles}
\author{J.-J. Ma$^{1}$, X.-X. Yuan$^{1}$, C. Zu$^{1}$, X.-Y. Chang$^{1}$,
P.-Y. Hou$^{1}$ L.-M. Duan$^{1,2}$}
\address{$^{1}$Center for Quantum Information, IIIS, Tsinghua University, Beijing 100084,
PR China}
\address{$^{2}$Department of Physics, University of Michigan, Ann Arbor, Michigan 48109, USA}
\date{\today}

\begin{abstract}
It was shown recently that entanglement of identical particles has a feature
called dualism [Phys. Rev. Lett. 110, 140404 (2013)], which is fundamentally
connected with quantum indistinguishability. Here we report an experiment
that observes the entanglement duality for the first time with two identical
photons, which manifest polarization entanglement when labeled by different
paths or path entanglement when labeled by polarization states. By adjusting
the mismatch in frequency or arrival time of the entangled photons, we tune
the photon indistinguishability from quantum to classical limit and observe
that the entanglement duality disappears under emergence of classical
distinguishability, confirming it as a characteristic feature of quantum
indistinguishable particles.
\end{abstract}

\pacs{03.65.Ud, 03.67.Ac, 03.65.Ta, 42.50.-p}
\maketitle
\section{Introduction}
Indistinguishability of identical particles is a fundamental feature of
quantum mechanics, which has deep consequence in quantum statistics and
many-body physics. The quantum indistinguishability has been confirmed for
various microscopic particles, ranging from the fundamental ones such as
photons \cite{1,2,3} or electrons \cite{4} to more complex composite objects
such as atoms \cite{5}. The test of quantum indistinguishability is usually
based on the Hanbury Brown-Twiss (HBT) type of experiment, which requires to
bring the particles together for high-order interference \cite{1,2,3,4,5}.
If the particles have mutual interaction with each other, which may become
unavoidable for increasingly massive objects, the interaction effect could
complicate the test of quantum indistinguishability. It was pointed out in a
very recent work that entanglement of identical particles shows a unique
property called duality, which is fundamentally connected with quantum
indistinguishability \cite{6}. This connection opens up a conceptually new
way to test quantum indistinguishability without the need of bringing the
particles together, thereby avoiding the interaction effect. The
entanglement duality means that if two identical particles are entangled in
a variable A when labeled by another variable B, they will also be entangled
in the variable B when labeled by the variable A. This feature is uniquely
associated with quantum indistinguishable particles and disappears when the
particles become distinguishable.

In this paper, we report the first experimental observation of the
entanglement duality with two identical photons and its fundamental
connection with quantum indistinguishability. The complementary variables A
and B are taken as the photon polarization and path. Through spontaneous
parametric down conversion in a nonlinear periodically-poled potassium
titanyl phosphate (PPKTP) crystal \cite{7,8}, we generate
frequency-degenerate photon pairs along two different paths labeled as
signal (S) and idler (I), which are entangled in polarization with an
entanglement fidelity of $\left( 98.5\pm 0.1\right) \%$. Then we separate
the photons according to their polarization label (horizontal or vertical)
and demonstrate their entanglement in the path variable (where the different
paths S and I are taken as the qubit basis states) with an entanglement
fidelity of $\left( 93.8\pm 0.3\right) \%$, thereby confirming the
entanglement duality for indistinguishable photons. To show that this
feature is uniquely associated with quantum indistinguishability, we make
the two photons distinguishable by adjusting the mismatch in their frequency
or arrival time. The mismatched frequency (or arrival time) is only correlated with the
path variable, so the initial symmetry between the two degrees of freedom (path and polarization)
is broken when we distinguish the photons by the new label of frequency (or arrival time).
As a result, in this case although we still observe a high amount of
entanglement in the polarization variable, we see no entanglement in the
path variable when the photons are separated according to their polarization.
\section{Theoretical background}
The concept of entanglement is defined for a composite system which can be
divided into two or more subsystems. Identical quantum particles are
indistinguishable when they are in the same state, so they can only be
labeled and separated through different states of certain variables (modes).
For instance, a photon can be labeled by different paths called signal (S)
or idler (I) or by different polarizations called horizontal (H) or vertical
(V), as illustrated in Fig. 1(a). To be concrete, let us consider a
polarization entangled state $\left\vert \Psi \right\rangle $ for two
photons between the signal and the idler modes with the following form%
\begin{equation}
\left\vert \Psi \right\rangle =\left( \left\vert H\right\rangle
_{S}\left\vert V\right\rangle _{I}+\left\vert V\right\rangle _{S}\left\vert
H\right\rangle _{I}\right) /\sqrt{2}.
\end{equation}%
Here, different states S and I of the path variable are used to label and
separate the two identical photons and different polarization states H and V
are taken as the qubit basis-vectors. The state $\left\vert \Psi
\right\rangle $ can be equally written into a dual form
\begin{equation}
\left\vert \Psi \right\rangle =\left( \left\vert S\right\rangle
_{H}\left\vert I\right\rangle _{V}+\left\vert I\right\rangle _{H}\left\vert
S\right\rangle _{V}\right) /\sqrt{2},
\end{equation}%
where we have switched the role of the path and the polarization variables,
with the photons entangled in the path variable while labeled by the
polarization variable. This is the entanglement duality first noticed in
Ref. \cite{6}, which holds only for identical particles. The dualism breaks
down when the particles along the paths S and I become distinguishable,
e.g., through difference in some other degrees of freedom such as frequency
or arrival time as illustrated in Fig. 1(b) and (1c). In this case, the
polarization entanglement remains the same and is still described by Eq.
(1). However, if we separate the photons according to their polarization,
the photons from the paths S and I are distinguishable in frequency (or
arrival time) and we therefore cannot observe any coherence or entanglement
in the path variable. Note that the concept of entanglement duality is
different from that of the hyper-entanglement \cite{9} although both of them
involve polarization and path entanglement. The hyper-entanglement is not
related to quantum indistinguishability and involves simultaneous
entanglement in polarization and path variables. Instead, the entanglement
duality is an intrinsic property of indistinguishable particles, where the
entanglement in polarization and path variables are complementary to each
other but not present simultaneously by the same measurement.

\begin{figure}[tbp]\centering
\includegraphics[width=8.5cm,height=7cm]{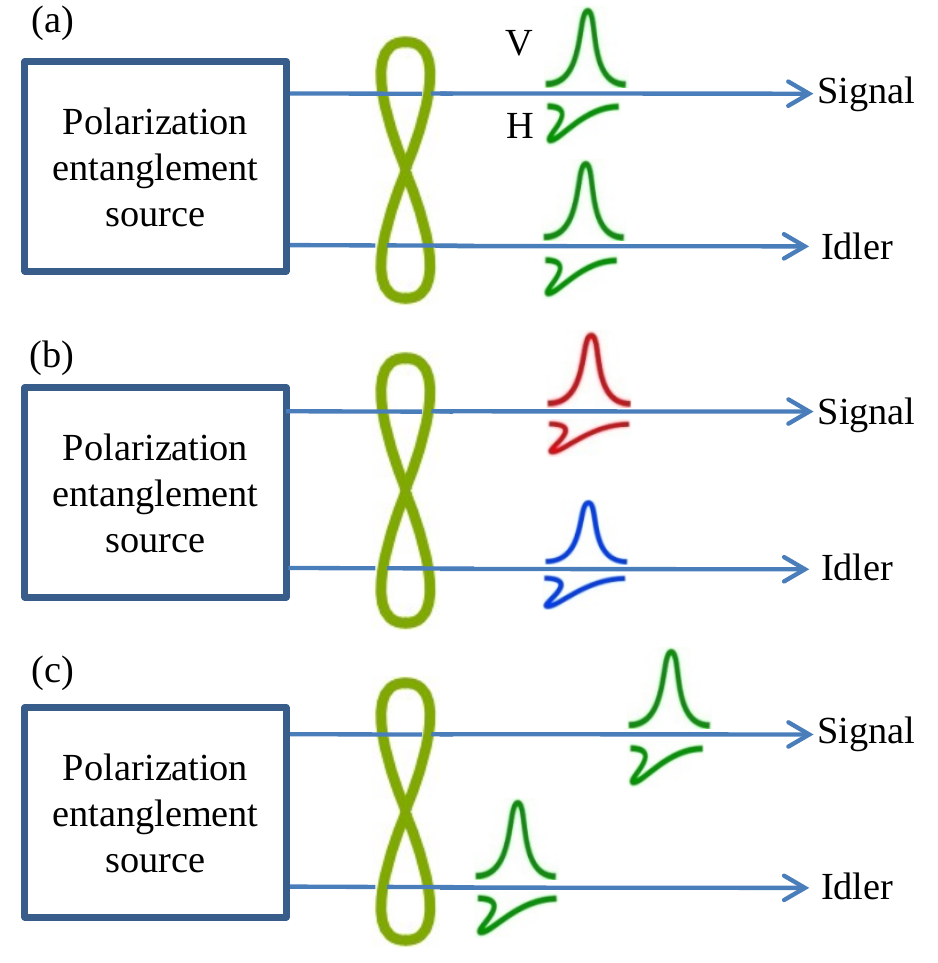}
\caption{Illustration of test of the entanglement duality through photons
with tunable quantum indistinguishability. (a) Entanglement for
indistinguishable photons which shows entanglement dualism in the
polarization (H and V) and the path (Signal and Idler) variables. (b,c)
Photons are distinguished through color (frequency, b) or arrival time at the detector (c)
and the entanglement dualism breaks down. }
\end{figure}
\section{Experimental facts}
To experimentally observe the entanglement duality, we first generate
frequency-degenerate photon pairs through spontaneous parametric down
conversion in a nonlinear PPKTP\ crystal. With the type-II phase matching in
the nonlinear crystal, the down converted photons have orthogonal linear
polarizations, denoted by $H$ and $V$, respectively. The entanglement is
produced through a sagnac interferometer as shown in Fig. 2 \cite{7,8}. When
the polarization of the pump beam is in a coherent superposition $\left(
\left\vert H\right\rangle _{P}+\left\vert V\right\rangle _{P}\right) /\sqrt{2%
}$, the down-converted photons go out of the interferometer along two paths
denoted as the signal and the idler modes, and the polarization of the
signal and the idler photons are described exactly by the entangled state
(1)\ in the ideal case.

%\begin{widetext}
\begin{figure}[tbp]\centering
\includegraphics[width=14.5cm,height=10cm]{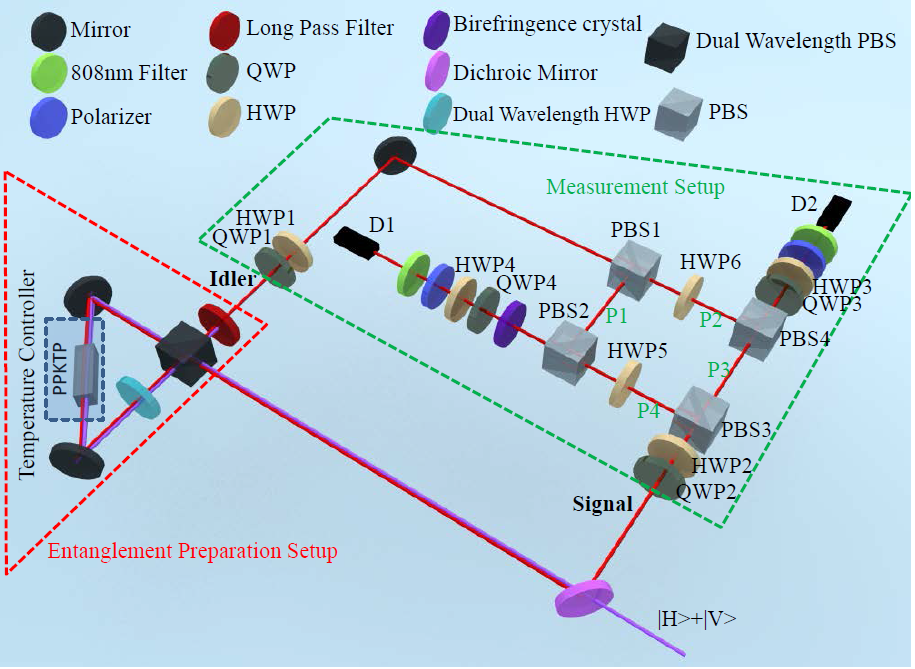}
\caption{Illustration of the experiment setup for test of the entanglement
duality. A continuous wave (cw) laser beam at the wavelength of $404$ nm and
with polarization $\left\vert H\right\rangle+\left\vert V\right\rangle$ from
a diode laser enters a sagnac interferometer to pump a PPKTP crystal ($15$
mm long with a cross-section $2\times1$ mm$^2$), generating down-converted
photon pairs at the wavelength $808$ nm through the type-II phase matching.
The interferometer is composed of a dual-wavelength (at both $404$ and $808$
nm) PBS (polarization beam splitter) and HWP (half wave plate, which flips
polarization between $H$ and $V$). The setup inside the red triangle
generates a polarization maximally entangled state between the signal and
the idler photons \protect\cite{7,8}. The setup inside the green box is for
measurement of the entanglement duality. By setting the angles of the HWPs
and QWPs at appropriate angles (see the text for details), the setup can be
used to measure either polarization (or path) entanglement when the photon
is labeled by path (polarization). A birefringent crystal is inserted after
PBS2 to compensate the optical length difference between the paths P1+P3 and
P2+P4. An interference filter of $3$ nm width centered at $808$ nm
wavelength is inserted before the single photon detectors D1 and D2, and the
photon counts of these detectors are registered through a home-made
coincidence circuit. This setup uses only single-photon interference to verify
entanglement in the path variable and never brings the two photons together for the
Hong-Ou-Mandel type of interference.}
\end{figure}
%\end{widetext}

In our experiment, we tune the photon distinguishability by adjusting the
mismatch in the frequency or the arrival time of the two photons. The
frequency mismatch between the two down-converted photons can be tuned
through adjustment of the temperature of the nonlinear crystal. At a given
temperature, the phase matching condition is satisfied only for certain
frequencies of the signal and the idler photons, and by tuning the
temperature, we can vary the frequency mismatch between the signal and the
idler photons. If this mismatch is larger than the bandwidth of the down
converted photons (which is about $0.78$ nm measured in term of the FWHM
(full width at the half maximum) of the wavelength spectrum, as illustrated
in Fig. 1(b), the photons in the signal and the idler modes are completely
distinguishable through the frequency. Alternatively, we can also
distinguish the signal and the idler photons through their different arrival
times at the photon detector. The down-converted photons have a large
bandwidth and thus a small coherence time about $2.8$ ps, which determines
the effective width of the temporal profile of the correlated photon pair.
If we tune the mismatch in the arrival time of the signal and the idler
photons to make it larger than the width of this temporal profile, the
photons are distinguishable through their different arrival times at the detector, as
illustrated in Fig. 1(c).

To confirm the entanglement duality for indistinguishable quantum particles,
we need to have a setup to measure the photon polarization entanglement when
they are labeled by the path and their path entanglement when the photons
are labeled by the polarization. The measurement setup shown in Fig. 2
achieve these two goals with the same apparatus. The setup consists of a
Mach-Zender interferometer composed by four polarization beam splitters
(PBSs) and a number of half-wave plates (HWPs) and quarter-wave plates
(QWPs). The detection of the polarization entanglement is straightforward.
We simply set the angles of HWP3, HWP4, HWP5, HWP6 and QWP3, QWP4 all at
zero so that they have no effect. In this case, PBS1 and PBS2 (PBS3 and
PBS4) work as polarizers to select out the vertical (horizontal)
polarization component for the idler (signal) photons, respectively. By
tuning the angles of HWP1 and QWP1 (HWP2 and QWP2) for the idler (signal)
photons, we can measure their polarization in arbitrary bases and then
reconstruct their polarization state through standard quantum state
tomography \cite{10}.

The main purpose of the measurement setup shown in Fig. 2 is for detection
of the path entanglement when the photons are labeled by their polarization
states. In this case, we set the angles of HWP1, HWP2, QWP1, QWP2 all at
zero. The PBS1 and PBS3 separate the idler and the signal photons according
to their polarization. Now we talk about entanglement between the horizontal
and the vertical photons. For the horizontal photon, the qubit basis-states
are denoted by the paths P2 and P3 (corresponding to the idler and the
signal modes, respectively). We need to measure the horizontal photon in P2,
P3, and their arbitrary superposition bases. For this purpose, we superpose
these two components at the PBS4 after the HWP6 set at angle $45^{o}$ (which
flips polarization from $H$ to $V$). Then, through rotation of the angles of
HWP3, QWP3, and the polarizer before the detector D2, we can measure the
horizontal photon in arbitrary superposition bases of the paths P2 and P3.
Similarly, through a combination of PBS2, HWP5 (set at angle $45^{o}$), and
rotation of HWP4, QWP4, we can measure the vertical photon in arbitrary
superposition bases of the paths P1 and P4. Note that for this measurement,
the two photons are never brought together for the HBT type of interference.
We detect each photon separately in their individual (superposition) bases.
From this measurement, we can reconstruct the path state through quantum
state tomography and derive its entanglement between the horizontal and the
vertical photons.

\begin{figure}[tbp]\centering
\includegraphics[width=15cm,height=9cm]{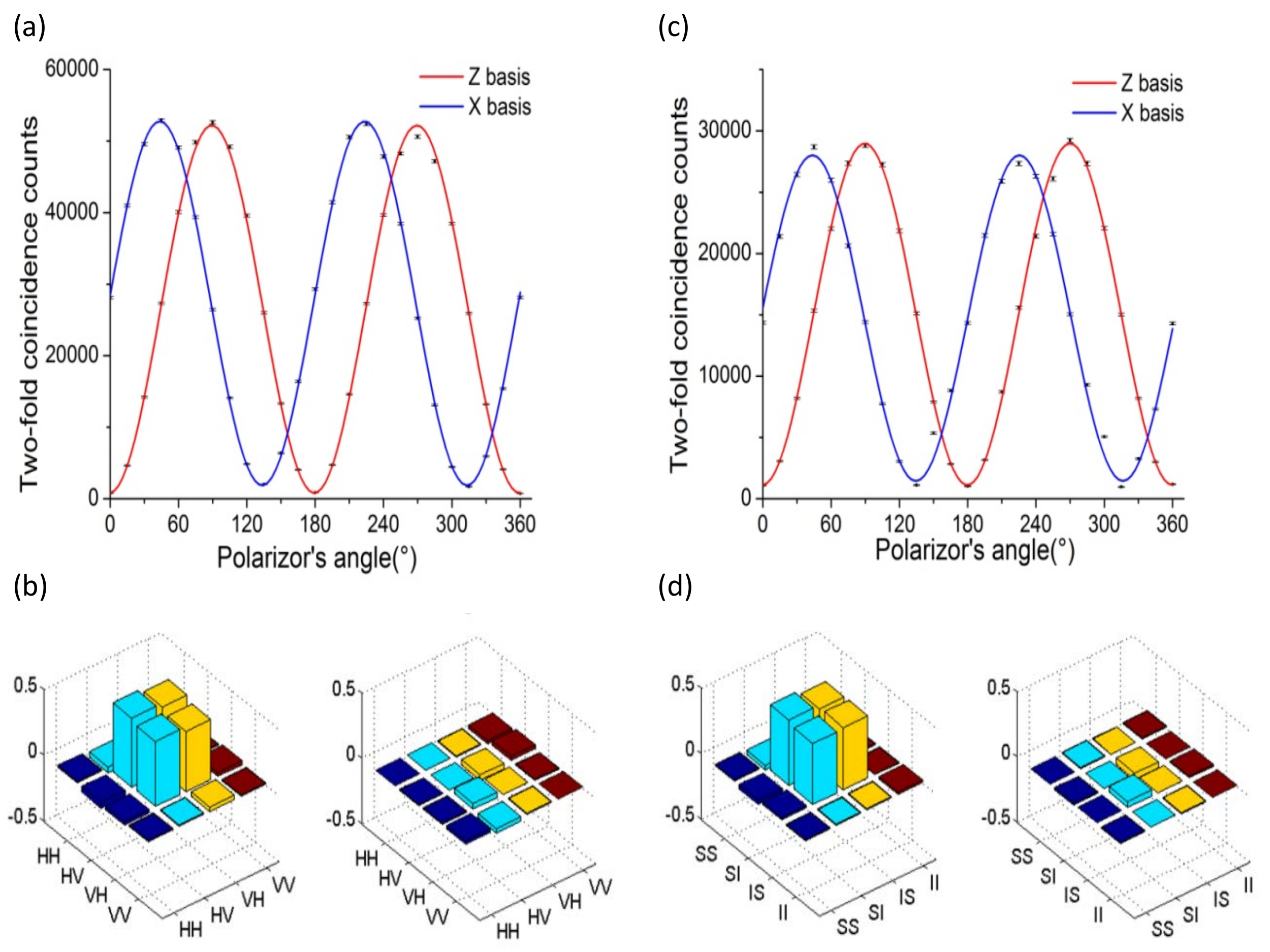}
\caption{Data for test of the entanglement duality with indistinguishable
photons, where Figs. a,b (c,d) show polarization (path) entanglement when
the photon is labeled by path (polarization). (a) The measured polarization
correlation in the complementary $Z$ and $X$ bases. (b) The reconstructed
density matrix for the polarization qubits, where the left (right) figure
shows the real (imaginary) parts of the matrix elements. (c) The path
correlation and (d) the density matrix of the path qubits for the horizontal
and the vertical photons.}
\end{figure}

The data from the entanglement duality measurement is shown in Fig. 3. In
Fig. 3(a) and 3(b), we show the data from measurement in the polarization
basis for the signal and the idler photons. First, by rotating the
polarizer's angle $\theta $ for the signal photon while fixing it to zero
for the idler photon, we measure their polarization correlation in the $Z$%
-basis (projection to $\left[ \cos \left( 2\theta \right) \left\vert
H\right\rangle _{S}+\sin \left( 2\theta \right) \left\vert V\right\rangle
_{S}\right] \left\vert H\right\rangle _{I}$) and the $X$-basis (projection
to $\left[ \cos \left( 2\theta \right) \left\vert H\right\rangle _{S}+\sin
\left( 2\theta \right) \left\vert V\right\rangle _{S}\right] \left\vert
+\right\rangle _{I}$ with $\left\vert \pm \right\rangle \equiv \lbrack \pm
\left\vert H\right\rangle +\left\vert V\right\rangle ]/\sqrt{2}$), and the
oscillation shown in Fig. 3(a) clearly demonstrates coherence of the input
state in the polarization basis. The full density matrix $\rho $ of these
two polarization qubits is reconstructed through quantum state tomography by
measurement of correlations in $16$ complementary bases \cite{10} and shown
in Fig. 3(b). From the measured density matrix, we calculate the
entanglement fidelity, defined as $F\equiv \left\langle \Psi \right\vert
\rho \left\vert \Psi \right\rangle $ ($\left\vert \Psi \right\rangle $ is
given by Eq. (1)) \cite{11}, and the concurrence $C$ defined in Ref. \cite%
{12} as a measure of its entanglement. We find $F=\left( 98.5\pm 0.1\right)
\%$ and $C=0.901\pm 0.002$, where the error bar accounts for the statistical
error associated with the photon detection assuming a Poissonian
distribution for the photon counts and the error bar is propagated through
exact numerical simulation. Similarly, in Fig. 3(c), we show the path
correlation in the $Z$-basis and the $X$-basis when the photons are labeled
by their polarization, and in Fig. 3(d), we show the measured density matrix
of the path qubits for the horizontal and the vertical photons. From the
measurement, we find $F=\left( 93.8\pm 0.3\right) \%$ and $C=0.896\pm 0.003$
for the path qubits. We therefore observe a high amount of entanglement for
the photons in the polarization variable when labeled by the path and in the
path variable when labeled by the polarization, which unambiguously confirms
the entanglement duality for two indistinguishable photons.

\begin{figure}[tbp]\centering
\includegraphics[width=15cm,height=10cm]{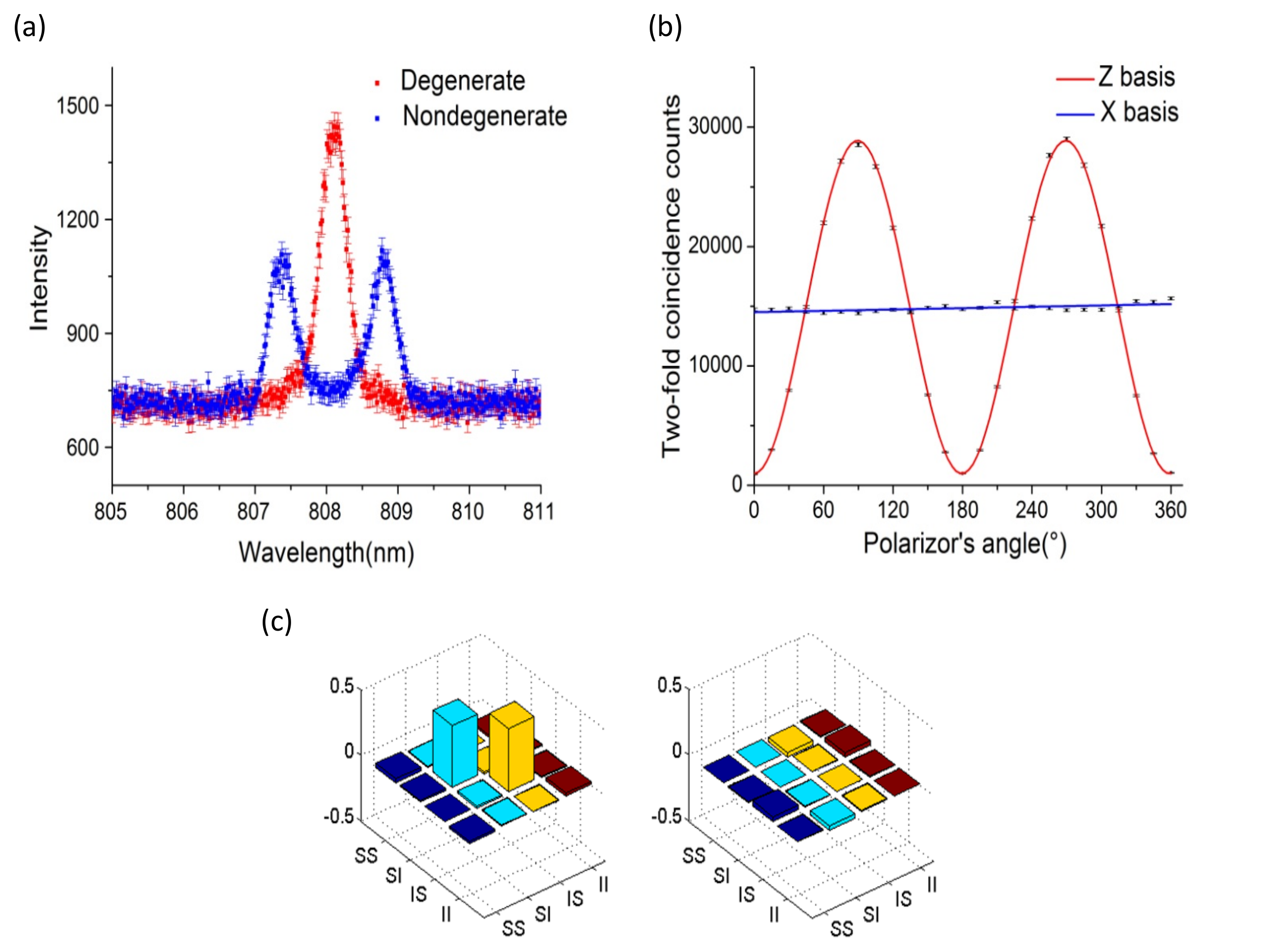}
\caption{(a) The spectrum of the down-converted photons measure through a
spectrometer, where the central peak shows the degenerate case with the
temperature of the PPKTP crystal is set at $53.7^{o}$ and the two edge peaks
correspond to the nondegenerate case with the crystal temperature at $%
50.0^{o}$. The photons are clearly distinguishable by frequency at the
non-degenerate case. (b,c) The measured path correlation (b) and the
reconstructed density matrix for the path qubits (c) when the photons are
distinguishable through either frequency or arrival time.}
\end{figure}

To demonstrate that the entanglement duality is connected with quantum
indistinguishability, we make the photons distinguishable in our experiment
by tuning up the mismatch in frequency or arrival time and observe the
corresponding change to quantum entanglement. As an example, in Fig. 4(a) we
show the measured frequency spectrum of the down converted photons by
adjusting the temperature of the PPKTP crystal. Clearly, we can tune the
photon distinguishability through this control knob. When the photons are
distinguished by color or arrival time, we observe a similar high amount of
entanglement in the polarization variable. For instance, with mismatched
frequencies as shown in Fig. 4(a), we find the concurrence $C=0.903\pm 0.004$
measured through the quantum state tomography. However, when we label the
photons by their polarization and measure their entanglement in the path
variable, we find no entanglement. As an example, we show in Fig. 4(b) and
4(c) the typical correlation curves and the density matrix elements for the
path variable when the photons become distinguishable by either color or
arrival time. The coherence is gone as indicated by the flat correlation
curve in the $X$-basis and the vanishing off-diagonal terms in the
reconstructed density matrix. From the measured density matrix, we find the
concurrence $C=0$, confirming no entanglement in the path variable. We
therefore demonstrate that the entanglement duality is a characteristic
property of indistinguishable particles and breaks down when particles
become distinguishable. Note that the entanglement (measured by the
concurrence $C$) in the dual basis is monotonically connected with quantum
indistinguishability and can be used as a quantitate indicator of the
latter, which attains the maximum value $1$ under perfect quantum
indistinguishability and the minimum $0$ for completely distinguishable
particles. In our experiment, the value of this quantity $C$ is $0.896\pm
0.003$ ($0)$ for the indistinguishable (distinguishable) particles.
\section{Conclusion}
In summary, we have reported the first experimental demonstration of
entanglement duality with two identical photons and its fundamental
connection with quantum indistinguishability. The experimental observation
of the entanglement duality offers a conceptually new way to test quantum
indistinguishability without the need of bringing the particles together for
the HBT type of interference.
\section*{Acknowledgement}
This work was supported by the National Basic Research Program of China (973
Program) 2011CBA00300 (2011CBA00302) and the NSFC Grants
61033001,61361136003. LMD acknowledges in addition support from the IARPA
MUSIQC program, the ARO and the AFOSR MURI program.

\end{document}